\documentclass[aps, prb, reprint, twocolumn, groupdaddress]{revtex4-1} 
\usepackage{graphicx}

\usepackage[usenames]{color}
\usepackage{colortbl}

\newcommand{\TV}[1]{}

\begin{document}

\title{Attosecond control of electrons emitted from a nanoscale metal tip}

\author{Michael Kr{\"u}ger}
\altaffiliation{These authors contributed equally to this work.}
\affiliation{Max Planck Research Group ``Ultrafast Quantum Optics'' \\Max-Planck-Institut f\"ur Quantenoptik \\\small Hans-Kopfermann-Str.~1, D-85748 Garching bei M\"unchen, Germany}
\author{Markus Schenk}
\altaffiliation{These authors contributed equally to this work.}
\affiliation{Max Planck Research Group ``Ultrafast Quantum Optics'' \\Max-Planck-Institut f\"ur Quantenoptik \\\small Hans-Kopfermann-Str.~1, D-85748 Garching bei M\"unchen, Germany}
\author{Peter Hommelhoff}
\affiliation{Max Planck Research Group ``Ultrafast Quantum Optics'' \\Max-Planck-Institut f\"ur Quantenoptik \\\small Hans-Kopfermann-Str.~1, D-85748 Garching bei M\"unchen, Germany}

\date{\today}

\maketitle

{\bf  Attosecond science is based on steering of electrons with the electric field of well-controlled femtosecond laser pulses~\cite{Corkum2007}. It has led to, for example, the generation of XUV light pulses~\cite{Antoine1996} with a duration in the sub-100-attosecond regime~\cite{Goulielmakis2008}, to the measurement of intra-molecular dynamics by diffraction of an electron taken from the molecule under scrutiny~\cite{Niikura2002, Baker2006}, and to novel ultrafast electron holography~\cite{Huismans2011}. All these effects have been observed with atoms or molecules in the gas phase. Although predicted to occur~\cite{Lemell2003, Stockman2007}, a strong light-phase sensitivity of electrons liberated by few-cycle laser pulses from solids has hitherto been elusive. Here we show a carrier-envelope \mbox{(C-E)} phase-dependent current modulation of up to $\bf{100\,\%}$ recorded in spectra of electrons laser-emitted from a nanometric tungsten tip. Controlled by the \mbox{C-E} phase, electrons originate from either one or two sub-500\,as long instances within the 6-fs laser pulse, leading to the presence or absence of spectral interference. We also show that coherent elastic re-scattering of liberated electrons takes place at the metal surface. Due to field enhancement at the tip, a simple laser oscillator suffices to reach the required peak electric field strengths, allowing attosecond science experiments to be performed at the 100-Megahertz repetition rate level and rendering complex amplified laser systems dispensable. Practically, this work represents a simple, exquisitely sensitive \mbox{C-E} phase sensor device, which can be shrunk in volume down to $\sim$1\,cm$^3$. The results indicate that the above-mentioned novel attosecond science techniques developed with and for atoms and molecules can also be employed with solids. In particular, we foresee sub-femtosecond \mbox{(sub-)} nanometre probing of (collective) electron dynamics, such as plasmon polaritons~\cite{Stockman2004}, in solid-state systems ranging in size from mesoscopic solids via clusters to single protruding atoms.}

A nanoscale solid-state system is of interest also for a more applied reason. The steering of electrons with the force exerted by a synthesized few-cycle light field is predicted to allow reaching ultimate speeds in electronics, i.e. up to optical frequencies with a typical time scale of femtoseconds (``lightwave electronics''~\cite{Goulielmakis2007}, in analogy to microwave electronics with semiconductor chips). However, typical electron energies in conventional electronics lie in the few electron volt range, corresponding to a velocity of a free electron of $\sim$1\,nm$/$fs. Thus, the speed-up of the drive can only be utilized in conventional electronics' complexity if a {\em nanometre-scale solid-state} system is employed~\cite{Stockman2007}. Because the electron current is switched on and off by the light field, one might call such a device an optical attosecond field-effect transistor. This work constitutes first research along these lines.

\begin{figure}
\begin{center}
\includegraphics[width=\columnwidth]{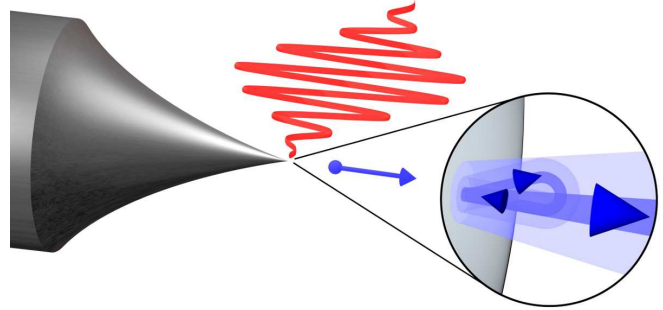}
\end{center}
	\caption{{\bf Overview of the experiment.}  Electrons, indicated in blue, are emitted from a sharp tungsten tip by, and interact with, a few-cycle laser electric field. Controlled by the carrier-envelope phase the liberated electron may be driven back into the tip, where it can scatter elastically and gain more energy in the laser field before it reaches the detector (not shown).}
	\label{fig:sketch}
\end{figure}

The experiment centers on a $\sim${}$10 \dots 20$\,nm-radius metal tip, whose apex lies in the focus of a linearly polarized laser beam consisting of few-cycle laser pulses, see Fig.~\ref{fig:sketch}. The tip is instrumental for two reasons. First, because of its sharpness electric field enhancement takes place~\cite{Novotny2011}. At the tip apex, the electric field is around five times higher than in the bare laser focus, corresponding to a $\sim$25-fold increase in intensity. Therefore, a laser {\em oscillator} suffices to reach the regime where Newton's equations of motion describe the electron's response quite well, marking attosecond science~\cite{Corkum2007}. Field-enhancement near nano-structures has also facilitated the generation of high-harmonic radiation from gas atoms with an oscillator only~\cite{Kim2008}, whereas in conventional gas phase experiments amplified laser systems are employed~\cite{Bucksbaum2007}. Second, because of the localized nature of field enhancement, electron emission is limited to a single well-defined site with a diameter of $\sim$10\,nm right on the tip's apex. As this is much smaller than the focal diameter of the laser beam, the laser intensity is well approximated as constant over the electron emission area. Thus electrons are emitted from a single nanometric area exposed to a well-defined laser intensity.

This does usually not hold for photoemission from plane surfaces, where local laser intensities greatly differ due to the Gaussian profile of the laser beam and hot-spot effects due to a contingent surface roughness. We conjecture that \mbox{C-E} phase effects are therefore blurred when the electron emission current from a larger area is measured. For presumably these reasons, only minuscule \mbox{C-E} phase effects have so far been observed in nonlinear photoemission from a gold cathode, even though record-short near-infrared laser pulses with a duration of 4\,fs were used~\cite{Apolonski2004}.  A previous attempt to measure a \mbox{C-E} phase dependence in photoemission from sharp tips lacked spectral information, which presumably hid the signal~\cite{Hommelhoff_PRL2006_2}. An attosecond streaking experiment from a solid has been reported in \bibpunct{}{}{,}{n}{}{}Ref.~\cite{Cavalieri2007}, but there the electrons were photo-emitted from the metal surface by XUV pulses before they interacted with the infrared light field, thereby mitigating hot-spot effects. Collective electron and strong-field effects have been observed in tip-enhanced electron emission in Ref.~\cite{Yanagisawa2009} and Refs.~\cite{Bormann2010} and~\cite{Schenk2010}, respectively, while Ref.~\cite{Kealhofer2011} shows that thermal effects can be ruled out here.

In the experiment we focus $\sim$6-fs pulses from a \mbox{C-E} phase stabilized, 80-MHz repetition rate Ti:sapphire oscillator tightly on a tungsten tip (Fig.~\ref{fig:sketch}. The setup is described in more detail in Ref.~\cite{Schenk2010} and in the Supplementary Information). A small negative extraction voltage is applied to the tip, resulting in a d.c. electric field strength of $\sim$0.4\,GV$/$m at the tip's apex. We record photoelectron spectra with a retarding field spectrometer.\bibpunct{}{}{,}{s}{}{}

\begin{figure*}
\includegraphics[width=118mm]{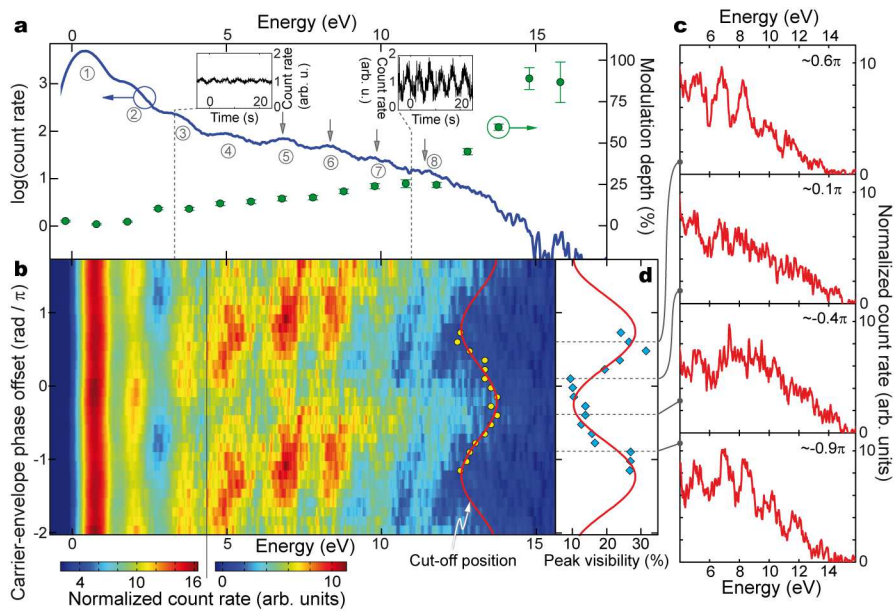} 
	\caption{{\bf Carrier-envelope phase modulation in photoelectron spectra.}  {\bf a}, Carrier-envelope phase-averaged electron count rate as function of energy (blue solid curve). About three photon orders, indicated by encircled numbers, are visible in the direct part ($E = 0 \dots \sim 4.3\,$eV). For $E > 4.5$\,eV the plateau region starts with five more photon orders visible. The green points depict the modulation depth of the count rate when varying the \mbox{C-E} phase (error bars: standard fit errors of modulation curves, see Supplementary Information). The insets show the \mbox{C-E} phase modulation in the photocurrent with the spectrometer acting as energy high-pass filter at 3\,eV and 11\,eV (with the \mbox{C-E} offset frequency set to $f_{\mathrm{CEO}} = \dot{\phi}_{\mathrm{CE}}/ 2\pi = 0.2\,\mathrm{Hz}$). {\bf b}, Contour plot of the electron count rate as function of \mbox{C-E} phase offset and energy. $\phi_{\mathrm{CE}}$ was increased in steps of $\pi /8$. At $4.3$\,eV the plot is split into two regions for better visibility. In each region we plot the normalized count rate (see Supplementary Information). Measured data range over $2\pi$ and are extended over $\sim 4\pi$ for better visibility. The yellow circles show the position of the cut-off for a given carrier-envelope phase offset (red curve: sinusoidal fit). {\bf c}, Individual electron spectra extracted from the contour plot in {\bf b} for four different carrier-envelope phase offsets separated by $\pi /2$. Only the plateau region is shown. Clearly, fringes are visible for $0.6\,\pi$ and $-0.9\,\pi$, but almost no fringes are visible for $0.1\,\pi$ and $-0.4\,\pi$. {\bf d}, Average peak visibility in the plateau region (red curve: sinusoidal fit). The peaks used to determine the visibility are marked with grey arrows in {\bf a}. Note that peak visibility and cut-off-position are nearly maximally out of phase.}
	\label{fig:ExpSpectra}
\end{figure*}

Already from the overall shape of the spectra conclusions can be drawn about the dynamics of the photoelectrons. In Fig.~\ref{fig:ExpSpectra}a we present a \mbox{C-E} phase averaged spectrum obtained with 240\,pJ pulse energy (peak intensity of $\sim 4\times 10^{11} \,$W$/$cm$^2$ in the bare focus), with an average yield of about one electron per pulse. The spectrum is governed by above-threshold photoemission (ATP) peaks approximately spaced by the photon energy (1.56\,eV) on top of an overall exponential decay~\cite{Schenk2010}. This decay is followed by a plateau, a region of almost constant count rate extending from $\sim$4.5\,eV to a soft cut-off located at $\sim$13\,eV. The appearance of the plateau indicates that coherent elastic re-scattering of electrons takes place, an effect well known in ionization of atoms in the gas phase: A small fraction of the photoelectrons is driven back into the tip by the laser field, scatters elastically off the tip, and gains more energy in the laser field before being detected~\cite{Paulus1994_PRL}. Theory models presented below strongly support this notion. A more detailed investigation of electron re-scattering dynamics from a metal is currently under way. Note that recombination of the active electron can lead to emission of high-harmonic radiation~\cite{Corkum2007}. We expect this process to also take place at tips.

The electric field of the laser pulses can be written as $E(t) = f(t) \cos (\omega t + \phi_{\mathrm{CE}}),$ with $f(t)$ describing the pulse envelope, $\omega$ the laser's center (circular) frequency, and $\phi_{\mathrm{CE}}$ the \mbox{C-E} phase. Fig.~\ref{fig:ExpSpectra}b shows a contour plot of individual electron spectra as a function of the \mbox{C-E} phase offset, which is given by the sum of $\phi_{\mathrm{CE}}$ and a constant experimental phase difference to be determined by theory (set to 0 here, see later). Clearly, the spectral features are strongly modulated with the \mbox{C-E} phase: Both maxima and minima show pronounced modulation effects with a period of $2\pi$. In Fig.~\ref{fig:ExpSpectra}a (green points) we display the modulation depth of the count rate for different energy positions (for a definition see Supplementary Information). At low energy the modulation depth amounts to several percent and gains strength in the plateau ($10 \dots 25\,\%$). In the region of the 13\,eV-cut-off it increases up to $\sim$100\,$\%$. Thus, here the \mbox{C-E} phase almost perfectly determines if a photoelectron is detected.

The visibility of the plateau peaks is particularly strongly affected by the \mbox{C-E} phase. Fig.~\ref{fig:ExpSpectra}c depicts individual spectra for four \mbox{C-E} phase settings spaced by $\pi /2$. It is evident that for certain \mbox{C-E} phases peaks are clearly visible, whereas for others the peaks almost fully disappear. An analysis of the average peak visibility is shown in Fig.~\ref{fig:ExpSpectra}d (for details see Supplementary Information). It is approximately sinusoidally modulated with the \mbox{C-E} phase, ranging from $\sim$$10\,\%$ for $\phi_{\mathrm{CE}} \approx -0.2\pi$ to $\sim$$30\,\%$ for $\phi_{\mathrm{CE}} \approx 0.8\pi$. We will show below that the peaks arise from quantum mechanical interference of electron wavepackets re-scattering at the tip in different optical cycles. The visibility can be identified as the degree of spectral interference. Strong interference indicates that (at least) two wavepacket components contribute to the plateau. In contrast, the absence of interference implies that only a single electron wavepacket from one optical cycle contributes.

In addition, we observe that the position of the high-energy cut-off changes with \mbox{C-E} phase as shown in Fig.~\ref{fig:ExpSpectra}b (red line). It varies between $12.3\,$eV and $13.6\,$eV. Notably, the behaviour of peak visibility and cut-off position is maximally out of phase: the phase-difference amounts to $\pi + (80 \pm 160)\,$mrad.

\begin{figure}
\includegraphics[width=\columnwidth]{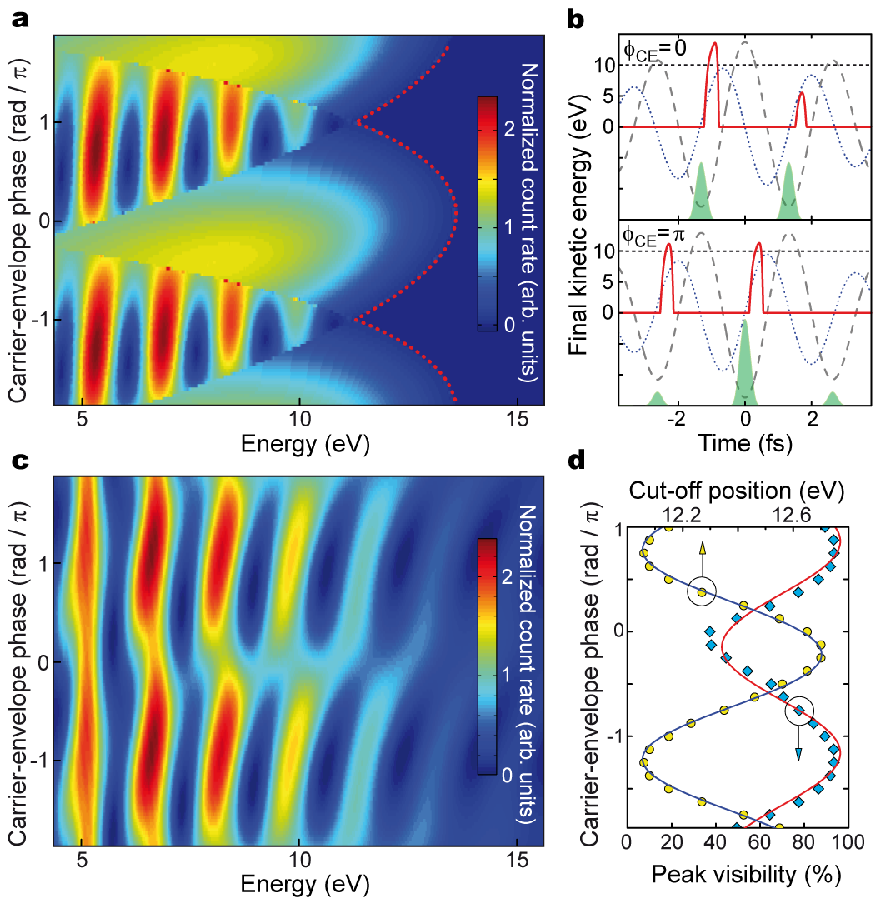}
	\caption{{\bf Theoretical modelling results of the experimental data.} {\bf a},
Contour plot representation of the count rate as a function of energy and carrier-envelope phase according to the extended Simple Man's Model. Here only re-scattered electrons are considered (only those contribute to plateau energies). Regions with quantum mechanical interference fringes ($\phi_{\mathrm{CE}} \approx \pm \pi$) and without ($\phi_{\mathrm{CE}} \approx 0$) are evident (the sharp borders reflect the semiclassical nature of the model). Furthermore, the highest electron energies are reached for \mbox{C-E} phase settings corresponding to about the most featureless electron spectrum ($\phi_{\mathrm{CE}} \approx 0$). The dotted red curve indicates the classical cut-off position. {\bf b}, Final kinetic energy (red curve) and emission probability (green area) as a function of emission time for phases $0$ and $\pi$. The emission probability in the model follows the electric field (dashed grey curve). Emission only occurs when the electric field is negative and points into the metal, thus pulling electrons out of the metal. The electron's kinematics can be inferred more easily from the light field's vector potential $A(t)$, shown with a blue dotted line ($E(t) = - \partial A(t) / \partial t$).  Notably, either one or two trajectories significantly contribute to the kinetic energy of, e.g., $10\,\mathrm{eV}$ (marked by the dotted horizontal lines). Within each cycle also a second trajectory exists, but the corresponding emission probability can be neglected against the dominating one. {\bf c}, The same as {\bf a}, but calculated by numerically integrating the time-dependent Schr{\" o}dinger equation. {\bf d}, Average peak visibility and cut-off position from {\bf c}.}
	\label{fig:TheorySpectra}
\end{figure}

We interpret our experimental findings with the aid of two theoretical models. The first model employed is the semiclassical Simple Man's Model (SMM)~\cite{Corkum_3Step_1993}. In brief, an electron is liberated by optically induced tunnelling and subsequently propagates in the laser electric field on classical trajectories. The model has been extended to account for the matter wave nature of liberated electrons by including the accumulated quantum mechanical phase of the corresponding wavepackets~\cite{MilosevicBecker2006}. Trajectories with different start times within the pulse that lead to the same final energy interfere, resulting in interference structures in the energy domain~\cite{Lindner2005}.

In Fig.~\ref{fig:TheorySpectra}a we present SMM spectra similar to Fig.~\ref{fig:ExpSpectra}b assuming a 6.3-fs pulse with a peak electric field of $10.4\,\mathrm{GV/m}$. All main features of the experimental data are qualitatively reproduced, notably the shift of the cut-off position and the correlated change in peak visibility. A region is observed where no spectral interference occurs, centered around $\phi_{\mathrm{CE}} \approx 0$ corresponding to a ``cosine-like'' pulse. Simultaneously, the cut-off position is located at the highest energy. Fig.~\ref{fig:TheorySpectra}b illustrates the physical origin of both effects for $\phi_{\mathrm{CE}} \approx 0$: Only a single trajectory from one optical half-cycle, reaching the highest possible kinetic energy, yields a significant contribution. In contrast, the region around $\phi_{\mathrm{CE}} \approx \pm \pi$, which corresponds to a ``minus-cosine-like'' pulse, exhibits the strongest peak structure and the lowest cut-off energy. Two trajectories from subsequent cycles contribute here with the time instants of both emission and rescattering differing approximately by the duration of an optical cycle $T_{\mathrm{opt}}$. This translates into interference fringes in the energy domain with a spacing of $\Delta E \sim h / T_{\mathrm{opt}} = h / 2.67\,\mathrm{fs} \approx 1.56$\,eV, about the photon energy~\cite{MilosevicBecker2006}. A similar behaviour was demonstrated for strong-field photoemission from atomic gases~\cite{Lindner2005}\bibpunct{}{}{,}{n}{}{}, and also in Ref.~\cite{Baltuska2003} where\bibpunct{}{}{,}{s}{}{} the generation of single sub-femtosecond XUV radiation bursts was deduced from the absence of interference in the cut-off region of the XUV spectrum. Accordingly, single electron matter wavepackets with initially sub-femtosecond emission duration (Fourier limit of cut-off part $\sim$450\,as) can be generated from a metallic tip by appropriately setting the \mbox{C-E} phase. Moreover, the timing of emission and propagation of electron wavepackets undergoing re-scattering near the metal surface can be controlled with a precision of $\sim$80\,as by changing the \mbox{C-E} phase, as inferred from the fit error of the phase offset in the sinusoidal fits to cut-off and visibility. The presence of interference indicates the coherent nature of the re-scattering process off a metal surface, which has not been reported before.

The second model, a fully quantum mechanical treatment, shows decent quantitative agreement with the experiment. In Fig.~\ref{fig:TheorySpectra}c we present energy spectra retrieved from a numerical integration of the one-dimensional time-dependent Schr{\"o}dinger equation (TDSE)~\cite{Hommelhoff_PRL2006_2}. The parameters were adapted to match the experimental data (see Supplementary Information for details). The average peak visibility and the cut-off position are depicted in Fig.~\ref{fig:TheorySpectra}d and behave similarly compared to the experimental data. We have shifted the \mbox{C-E} phase axis of the experimental data so that the maxima of the cut-off position curves coincide, thereby zeroing the experimental phase offset. Thus, the pulse is cosine-like ($\phi_{\mathrm{CE}} = 0$) for the \mbox{C-E} phase offset of $(0.00 \pm 0.05) \pi$. The spectral shift of the cut-off with \mbox{C-E} phase in the TDSE model is a weaker effect (peak-to-peak $\sim 0.7\,$eV) than in the experimental data (peak-to-peak $\sim 1.3\,$eV, cf. Fig.~\ref{fig:TheorySpectra}d and Fig.~\ref{fig:ExpSpectra}b) mainly because strong smoothing had to be applied to the TDSE spectra in order to suppress effects caused by the pronounced peak structure. The SMM model (Fig.~\ref{fig:TheorySpectra}a), in contrast, reveals a larger shift (peak-to-peak $\sim 2.3\,$eV) mainly due to its semiclassical nature. Given the simplicity of the models, the agreement is satisfactory.

The peak visibility is lower in the experiment than in both theoretical treatments, which we attribute to the spectrometer resolution and low counting statistics. Moreover, in both models we consider only a single initial electronic state at the Fermi level. A metal, however, comprises many populated states with a large spread of initial energies. In future work, a more elaborate simulation~\cite{Lemell2003, Stockman2007} is needed to fully take this as well as possible plasmonic effects into account and to use the technique to draw conclusions about the underlying sub-femtosecond dynamics.

Three points are noteworthy. The observable \mbox{C-E} phase of the enhanced field at the tip's surface should be phase-shifted with respect to the phase of the driving laser field due to the plasmonic response of the metal~\cite{Stockman2007, Kim2008}. Spectra measured from a tip with strong plasmonic behaviour such as a silver tip will thus allow obtaining information of the collective electron response on the (sub-) nanometre--sub-femtosecond scale. Second, in the SMM an emission process according to optically induced tunnelling has been assumed, with the electrons' classical trajectories starting at the tunnel exit with zero initial momentum~\cite{Huismans2011,Zherebtsov2011} and with an emission probability modelled along the lines of the ADK rate (see Supplementary Information). We also tried other emission processes (non-adiabatic tunnelling~\cite{Yudin2001}, multiphoton photoemission) but found the best agreement using tunnelling, although here the Keldysh parameter is $\sim$2. We stress that in this parameter range the emission process encompasses and cannot be separated from strong-field effects after the electron has been born classically. Third, while the comprehensive understanding of the exact quantum dynamics of electrons in this new system is complex and requires much further investigation, it has become clear from experiments with neutral atoms~\cite{Paulus1994_PRL} and negatively charged ions~\cite{Gazibegovic-Busuladzic2010} that the re-scattering scenario and the concomitant telltale plateau seem to be universal in the sense that they exist despite qualitatively different potentials (long~\cite{Paulus1994_PRL} vs. short range~\cite{Gazibegovic-Busuladzic2010}). First theoretical work on photoemission from metals also hints in this direction~\cite{Faisal2005}. Our experimental results bear strong evidence for very similar underlying physics, even though here the dimensions involved are wholly different than in the atom or ion case: The electron source and scatterer, namely the solid tip, is much larger than the classical oscillation amplitude of the electron in the laser field (a few angstrom). It will be interesting to investigate what the implications are. For example, does scattering take place at the extended surface or at individual surface atoms? Does this depend on the material and its orientation? Angle-resolved spectra might yield information.

With longer laser wavelengths, the energy of the re-colliding electron increases and can surpass several ten electronvolts. Hence, new forms of (time-resolved) surface science techniques such as low-energy electron diffraction (LEED), with electrons originating from and probing the surface, might come into reach, with typical time scales of 100\,as. Extending this work towards more complex, lithographically grown nano-objects~\cite{Novotny2011,Aeschlimann2007} will pave the way towards lightwave electronics, where the light electric field steers electrons and thereby, for example, can switch between the conducting or insulating state between a source and a drain structure. Also, a simple record-sensitive stand-alone die-sized sensor device for the \mbox{C-E} phase, comprising of a tip, a retardation grid, and an electron multiplier, might result.


\section*{Acknowledgements}
We thank M. Kling, C. Lemell, G. Wachter, and B. Bergues for discussions and J. \mbox{Hoffrogge} for proofreading. This work has been supported in part by the European Union (FP7-IRG).

\section*{Author Contributions}
All authors contributed to all parts of the work.

\section*{Author Information}
Correspondence and requests for materials should be addressed to P.H. (peter.hommelhoff@mpq.mpg.de).

\end{document}


\title{Attosecond control of electrons emitted from a nanoscale metal tip\\-- Supplementary Information --}

\author{Michael Kr{\"u}ger}
\altaffiliation{These authors contributed equally to this work.}
\affiliation{Max Planck Research Group ``Ultrafast Quantum Optics'' \\Max-Planck-Institut f\"ur Quantenoptik \\\small Hans-Kopfermann-Str.~1, D-85748 Garching bei M\"unchen, Germany}
\author{Markus Schenk}
\altaffiliation{These authors contributed equally to this work.}
\affiliation{Max Planck Research Group ``Ultrafast Quantum Optics'' \\Max-Planck-Institut f\"ur Quantenoptik \\\small Hans-Kopfermann-Str.~1, D-85748 Garching bei M\"unchen, Germany}
\author{Peter Hommelhoff}
\affiliation{Max Planck Research Group ``Ultrafast Quantum Optics'' \\Max-Planck-Institut f\"ur Quantenoptik \\\small Hans-Kopfermann-Str.~1, D-85748 Garching bei M\"unchen, Germany}



\setcounter{page}{6}

{
\section*{Supplementary Material}

\section{Experimental setup}

The basis of the experimental setup has been described elsewhere~\cite{Schenk2010}, hence we only give a brief summary here. The light source in our experiment is a Ti:sapphire oscillator, which provides near-infrared $\sim$6-fs pulses operating at a repetition rate of 80\,MHz. The carrier-envelope (\mbox{C-E}) phase of the oscillator pulses is stabilized to a constant value with the help of a $f$-to-$2f$ interferometer~\cite{Holzwarth2000, Jones2000} with an acousto-optic modulator in one arm~\cite{Jones2000, Hommelhoff2006IEEE}, using an octave spanning frequency comb generated with a photonic crystal fibre. We change the \mbox{C-E} phase by varying the phase of the reference signal. A second, out-of-loop $f$-to-$2f$ interferometer based on a periodically poled lithium niobate crystal~\cite{Fuji2005OL} serves for monitoring the long-term (minute scale) behaviour of the stabilization system. The mean phase drift of the \mbox{C-E} phase when phase-locked is below 100\,mrad$/$min as monitored with the out-of-loop $f$-to-$2f$ interferometer. During each measurement we correct for phase drifts with the help of this interferometer.

In an ultrahigh vacuum chamber \OK{(CVT Ltd.)} with a pressure of $\sim 3\times 10^{-8}$\,Pa, the laser beam  is tightly focused on the apex of a tungsten tip with a spot size of $\sim 2.4\,{\mathrm{\mu m}}$ ($1/e^2$ intensity radius). The polarization axis of the linearly polarized laser beam is parallel to the tip's pointing direction. The tungsten tip is fabricated from a W(310) single crystal wire by electrochemical etching and has a radius of curvature at the apex of about $10\dots 20$\,nm. The tip radius has been determined by two methods. {\em Ex-situ} imaging of the tip in a scanning electron microscope gives a upper bound of $30$\,nm. The ring counting method in an {\em in-situ} field ion microscope image~\cite{Tsong1990} yields a radius of about 10\,nm. We record photoelectron spectra with a retarding field spectrometer. A small negative extraction voltage (50V) is applied to the tip, resulting in a d.c.~electric field strength of $\sim$0.4\,GV$/$m at the tip's apex.

\section{Analysis of the experimental data}

\subsection{Data processing}

In order to obtain a single spectrum as displayed in Fig.~2c several data processing steps have to be taken. We record a series of 10 integrated spectra (count rate vs retardation voltage) with the retarding field spectrometer and average over them. The energy scan step size is 13\,meV with a time window of 5\,ms for counting photoelectron events for each energy. In order to obtain a spectrum we take the derivative of the count rate with respect to the energy and smooth the resulting curve with the Savitzky-Golay algorithm spanning 1.5\,eV. The energy axis in the measurement is calibrated to the Fermi level $E_{\mathrm{F}}$. In the data presented, however, the energy axis's origin was chosen to be the vacuum level, which is located $\sim$5.2\,eV above the Fermi level, as determined from the experiment. The energy axis then represents the kinetic energy of the photoelectrons. In a further processing step the whole series of spectra as a function of \mbox{C-E} phase is carefully smoothed along the \mbox{C-E} phase axis: A second-order Savitzky-Golay smoothing algorithm involves 5 neighbouring data points only, balancing successful smoothing with a possible loss of information. The algorithm was applied to the measurement data extended over $4\pi$ to prevent boundary effects from affecting the measured data. After processing we estimate the effective resolution of the data in the plateau region to be around 500\,meV. This value is larger than the resolution of the spectrometer ($\sim 80$\,meV) \OK{due to the applied smoothing, \OK{which is} necessary because of the low count rate in this part}. In order to show the \mbox{C-E} phase effects as clearly as possible in the plots of Fig.~2b and 2c we divided the count rate in each of the two regions (direct part and plateau) by an exponential decay curve approximating the respective shape. The result is the normalized count rate.

\subsection{Carrier-envelope phase modulation depth}

The modulation depth of the count rate for varying \mbox{C-E} phase at a given energy position as displayed in Fig.~2a is defined as $(N_{\mathrm{max}} - N_{\mathrm{min}}) / (N_{\mathrm{max}} + N_{\mathrm{min}})$ where $N_{\mathrm{max}}$ is the maximum and $N_{\mathrm{min}}$ the minimum count rate. We used the original spectral data, i.e., without dividing by an exponential decay curve, unlike in Fig. 2b and 2c. In Fig.~\ref{fig:ExpModDepth} we show individual modulation curves for two energies. In order to suppress noise we averaged over the count rate of a spectral region of width 1.5\,eV centered around each energy position. The count rate is almost fully modulated in the curve taken at an energy higher than the cut-off (see Fig.~\ref{fig:ExpModDepth}). It is also evident that the count rate reaches its maximum at different \mbox{C-E} phase offsets throughout the spectrum. The modulation depth was calculated from a sinusoidal fit to each modulation curve. The error bars in Fig.~2a represent the normalized amplitude error of this fit.

\begin{figure}
\includegraphics[width=0.83\columnwidth]{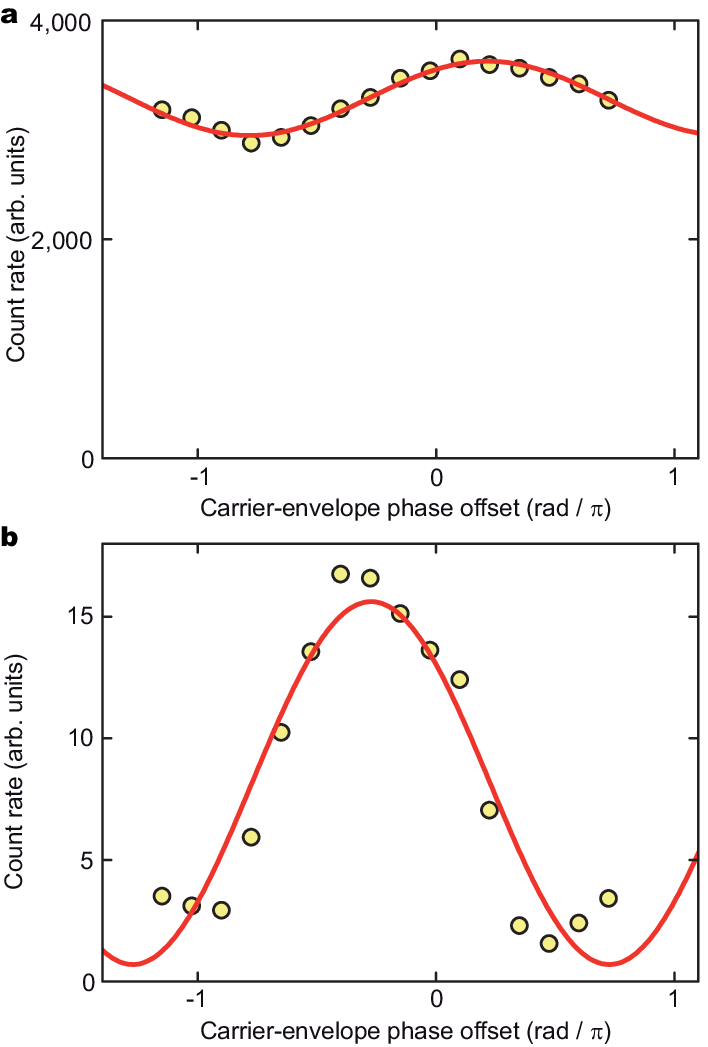}
\caption{{\bf Modulation in the energy spectrum for two energies.} {\bf a}, Modulation in the count rate as a function of \mbox{C-E} phase offset at  2.8\,eV (red curve: sinusoidal fit). {\bf b}, The same for 14.8\,eV.}
\label{fig:ExpModDepth}

\end{figure}

\subsection{Peak visibility}

In Fig.~2d the average peak visibility in the plateau is shown. The visibility of a single peak is defined as $(A - B) / (A + B)$ where $A$ is the count rate at the peak's maximum and $B$ is the average of the count rates of the two minima next to the peak. Also here we used the original non-normalized spectra. In Fig.~\ref{fig:ExpVisibility} we present an example analysis of four peaks in the plateau. If peaks are clearly visible (as in the case of Fig.~\ref{fig:ExpVisibility}) we performed a fit comprising of multiple Gaussian peaks and evaluated the visibility from the fit curve. In the case of weak peak structure we applied strong Savitzky-Golay smoothing in order to suppress noise obscuring the peaks.

\begin{figure}
\includegraphics[width=0.83\columnwidth]{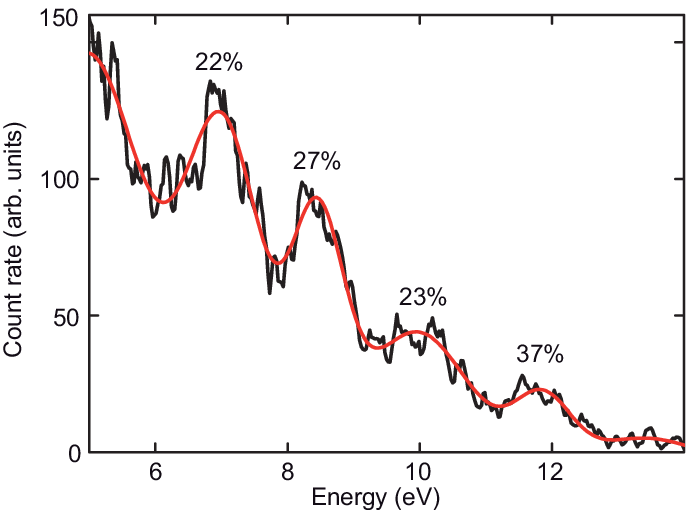}
\caption{{\bf Example analysis of the peak visibility in a photoelectron spectrum.} Count rate as a function of energy for a \mbox{C-E} phase offset of $-0.9\pi$. A fit with multiple Gaussian peaks was performed (red curve). The visibility of each peak was determined from this fit curve.}
\label{fig:ExpVisibility}
\end{figure}

\subsection{Cut-off position}

\begin{figure}
\includegraphics[width=\columnwidth]{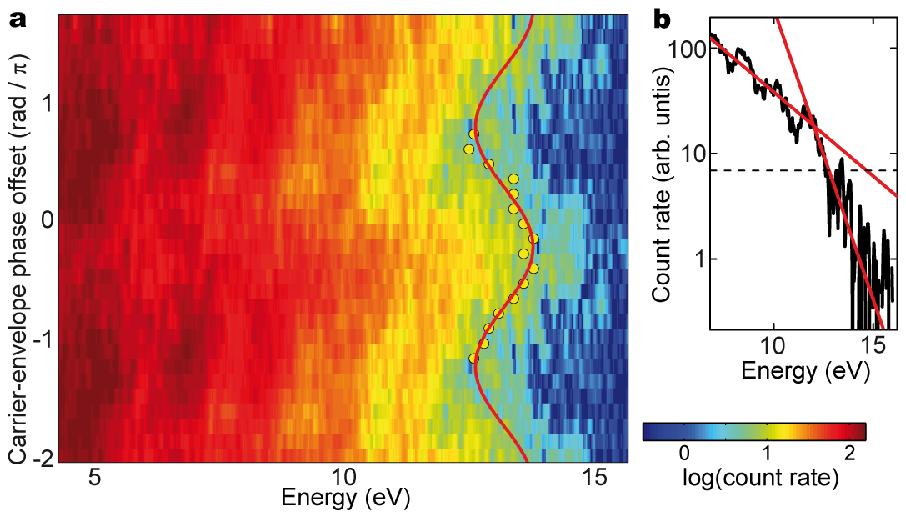}
\caption{{\bf Analysis of the cut-off position.} {\bf a}, Count rate in logarithmic scale as a function of energy and carrier-envelope phase offset. Only the plateau part is shown. Yellow circles mark the position of the cut-off as shown also in Fig.~2b (red: sinusoidal fit). The colour encoding of the contour plot suggests the behaviour of the cut-off equally well. {\bf b}, Count rate as a function of energy for the \mbox{C-E} phase offset $-0.9\pi$. Two different exponential slopes are fitted to the count rate nearby the cut-off. The cut-off position is defined as the intersection of the fit to the steeper slope with a horizontal line of constant count rate (7 arb. units). We found that taking the intersection of the exponential slopes as the cut-off position is not sufficiently robust against noise in the data.}
\label{fig:ExpCutOffLogPlot}
\end{figure}

The modulation of the cut-off position is directly visible in a contour plot similar to Fig. 2b, but with the count rate displayed on a logarithmic scale. In Fig.~\ref{fig:ExpCutOffLogPlot}a we present such a plot. The cut-off position has been quantitatively determined using exponential decay fits (see Fig.~\ref{fig:ExpCutOffLogPlot}b).

\section{Extended Simple Man's Model}

In the Simple Man's model~\cite{Corkum_3Step_1993} (SMM) the trajectory of an electron until its detection is split up in three steps. The first step, the emission of a photoelectron, is modelled in our embodiment of the SMM by an instantaneous tunnelling emission process. We chose an emission rate similar to the Ammosov-Delone-Krainov (ADK) rate~\cite{AmmosovDK1986}, which was formulated for the ionization of atomic gases. The emission rate $W$ as a function of time $t$ is given by
\begin{eqnarray*}
W(t) = \frac{A}{|E(t)|} \Theta [-E(t)] \exp{\left(- \frac{4\sqrt{2m}{\phi}^{3/2}}{3 \hbar |eE(t)|} \right)}.
\end{eqnarray*}
$A$ is a constant in time, $E(t)$ the laser electric field, $\Theta(x)$ the heaviside function, $m$ the electron mass, $e = -|e|$ the electron charge, $\hbar$ the reduced Planck constant and $\phi$ the work function of the metal. The electric field is defined as ${\bf E}(t) = {\bf e}_{\mathrm{z}} f(t) \cos (\omega t + \phi_{\mathrm{CE}})$. ${\bf e}_{\mathrm{z}}$ is the unit vector perpendicular to the metal-vacuum interface ($z$-axis), $\omega = 2\pi c / \lambda$ the (circular) frequency of the laser field corresponding to the central laser wavelength $\lambda$ and $c$ the vacuum speed of light. The pulse envelope $f(t)$ is modelled with a sine-square pulse, \OK{$f(t) = E_0 \sin^2[\omega t / (2n)]$} with the peak electric field $E_0$ and $n = \omega \tau / [4 \arccos(2^{-1/4})]$ carrier wave cycles covered by the envelope. $\tau$ is the full width at half maximum (FWHM) duration of the intensity envelope. \OK{The electric field $E(t) = {\bf E}(t)\cdot {\bf e}_{\mathrm{z}}$ is defined to be positive when the field points out of the metal into free space. The classical Lorentz force exerted by the laser field on an electron (negative point-like particle) is given by ${\bf F}(t) = -|e| {\bf E(t)}$. Thus an electron is pushed into the metal for positive fields. Contrary to atomic potentials the left-right symmetry in the direction of emission is broken at the metal surface. In the tunnelling picture, no emission occurs when the laser electric field is positive. This is accounted for by the heaviside function term in the formula for the emission rate.} For the SMM results presented in Fig. 3a and 3b we chose the experimentally determined work function of $\phi = 5.2$\,eV, a central wavelength of $\lambda = 800$\,nm, a pulse duration of $\tau = 6.3$\,fs and a peak electric field of $E_0 = 10.4\,\mathrm{GV/m}$. $E_0$ was set so that the appearance of the plateau qualitatively matches the \OK{experimental data} (cf. Fig.~2b and Fig.~3a). The field enhancement factor is about \MK{6}, inferred from the ratio of the chosen peak electric field and the experimentally expected value of \MK{$(1.8\pm 0.4)$\,GV/m} in the bare laser focus (without tip). With this model, a smaller factor of, for example, 4 as inferred from other analyses would lead to very different looking spectra. The ponderomotive energy in the chosen parameter set is $U_{\mathrm{p}} = 0.86$\,eV.

In the second step, a photoelectron emitted at an emission time $t_0$ is propagated classically in the laser field along the $z$-axis. The photoelectron starts with zero velocity at the position of the tunnel exit $z_{\mathrm{exit}}(t_0) = - \phi / (|e| E(t_0))$. The trajectory of the electron is evaluated numerically, and only trajectories are considered that lead to re-collision, i.e., that return to the metal-vacuum interface at $z=0$ and elastically scatter there at a time instant $t_1$. \OK{Note that direct electrons emitted without re-collision do not contribute to the plateau region in the SMM model because they only reach a maximum kinetic energy of $2U_\mathrm{p}$ (Ref.~\bibpunct{}{}{,}{n}{}{}\cite{MilosevicBecker2006}\bibpunct{}{}{,}{s}{}{}), here $\sim 1.7$\,eV. }All trajectories were ignored where the electron spends more than one optical cycle in the laser field upon re-collision at the surface.\OK{ Scattering is modelled by reflection off a hard wall with a probability of 100\%\TV{ that an electron scatters elastically}, independent of its incident kinetic energy.} We emphasize here that the initial position of the classical electron (tunnel exit $z_{\mathrm{exit}}$) and the position where re-scattering occurs (metal-vacuum interface) do not coincide, similar to\bibpunct{}{}{,}{n}{}{} Ref.~\cite{Zherebtsov2011}\bibpunct{}{}{,}{s}{}{} and in contrast to the original approach~\cite{Corkum_3Step_1993}. Such a treatment, however, resembles more closely a quantum mechanical treatment according to the Quantum Orbit Theory~\cite{Salieres2001} where a starting position at $z \neq 0$ close to $z_{\mathrm{exit}}$ can be identified~\cite{MilosevicBecker2006}. Note that this also leads to a correction of the ``$10 U_{\mathrm{p}}$ law'' for the cut-off position in the energy spectra to a higher value similar to the considerations in\bibpunct{}{}{,}{n}{}{} Ref.~\cite{Busuladzic2006}\bibpunct{}{}{,}{s}{}{}. In our case the cut-off is located at $\sim${}$16 U_{\mathrm{p}}$.

In the third step, we evaluate the final kinetic energy $E_{\mathrm{kin}}(t_0, t_1)$ at the detector for each trajectory starting at an emission time $t_0$. It is given by $E_{\mathrm{kin}}(t_0, t_1) = {p_1(t_0, t_1)}^2 / (2m)$ where $p_1(t_0, t_1)$ = $-|e| [2A(t_1) - A(t_0)]$ is the drift momentum of the re-scattered electron and $A(t)=-\int^t \mathrm{d}t' E(t')$ the vector potential. In order to account for the quantum mechanical nature of the electrons we assign a quantum mechanical phase $\theta(t_0, t_1)$ to each trajectory following\bibpunct{}{}{,}{n}{}{} Ref. \cite{PaulusLindner2004}\bibpunct{}{}{,}{s}{}{}. It is defined as $\theta(t_0, t_1) = S(t_0, t_1) / \hbar$ where $S(t_0, t_1)$ is the \OK{quasiclassical} action along the trajectory. The action formulated along the lines\bibpunct{}{}{,}{n}{}{} of Ref.~\cite{MilosevicBecker2006}\bibpunct{}{}{,}{s}{}{} is given by
\begin{eqnarray*}
S(t_0, t_1) = \int^{t_1} \mathrm{d}t \{ [p_1(t_0, t_1) + |e|A(t)]^2/(2m) + \phi \}\\  - \int_{t_0}^{t_1} \mathrm{d}t \{ [p_0(t_0) + |e|A(t)]^2/(2m) + \phi \}.
\end{eqnarray*}
$p_0(t_0) = - |e| A(t_0)$ is the drift momentum of the electron emitted at $t_0$ without considering re-scattering. \OK{The work function $\phi$ enters here in order to account for the quantum-mechanical evolution of the initial state. }The action $S(t_0, t_1)$ is evaluated numerically.\TV{of $\pi$ upon re-scattering is included in the calculation.} All contributions of trajectories to a discrete final energy are summed coherently and are weighted with the emission rate at the corresponding emission time. The probability $P(E_{\mathrm{kin}})$ to detect an electron at final energy $E_{\mathrm{kin}}$ is then given by
\begin{eqnarray*}
P(E_{\mathrm{kin}}) = \left| \sum_j \left\{ \sqrt{W(t_0^{(j)})} \exp[i \theta(t_0^{(j)}, t_1^{(j)})] \right\} \right|^2
\end{eqnarray*}
where $j$ numbers the contributing trajectories. The spectral interference pattern of two trajectories is dominated by an oscillation of the type $P(E_{\mathrm{kin}}) \propto B + C \cos(\Delta S / \hbar + \phi_0)$ where the action difference is given by $\Delta S = S_2 - S_1 \approx E_{\mathrm{kin}} \Delta t_1$ ~\cite{MilosevicBecker2006}. Here $B$ and $C$ are constants and $\phi_0$ a phase shift with a neglible dependence on the trajectories considered. If the difference $\Delta t_1$ of the time instants of re-scattering is approximately equal to the duration of one optical cycle $T_{\mathrm{opt}}$ a spectral interference pattern with a fringe spacing of around the photon energy is observed. This is the dominating effect for our parameter set. From Fig.~3b, however, it is evident that there is one more \OK{contributing trajectory} located in the same optical cycle\TV{,} but at a slightly later ($\sim$200\,as) emission time. Interference of these two intra-cycle trajectories would lead to spectral oscillations with a much larger spacing of around 20\,eV. This effect, however, is not discernable because the emission rate attributed to the secondary trajectory is negligibly small in comparison.\OK{ The value of the scattering phase shift upon re-scattering does not influence the resulting spectra and their interference structure because we only take electrons into account that re-collide. The scattering phase could only be measured if it led to a relative phase, but here it is a common phase to both scattering events leading to interference and thus an unmeasurable global phase.}

\section{Numerical integration of the Schr{\" o}dinger Equation}

The method used here has been described elsewhere~\cite{Hommelhoff_PRL2006_2, Hommelhoff2006IEEE}. In brief, the one-dimensional time-dependent Schr{\" o}dinger equation (TDSE) is numerically solved for a single active electron. The initial state is the ground state in a narrow potential well. The size of the well is chosen so that the ground state energy (measured from the bottom of the well) is matched to the Fermi energy of tungsten ($\sim$9\,eV). The metal-vacuum barrier is modelled as a potential step along with a correction given by the image force potential. The static electric field of magnitude $0.4\,\mathrm{GV/m}$ is also included in the calculation, in contrast to the SMM approach. It bends down the potential barrier and decreases its height by $\sim${}$0.8$\,eV due to the Schottky effect~\cite{Schottky1914}. We chose a work function of 6.0\,eV so that\OK{,} including the Schottky effect\OK{,} the effective barrier height is 5.2\,eV. The work function is significantly higher than the value of $4.35$\,eV reported for a tungsten surface in W(310) orientation~\cite{Kawano2008}. This could be explained by adsorption of atoms from the residual gas in the chamber~\cite{Yamamoto1978}. We regularly observe an increase of the effective barrier height within an hour after cleaning the tip, but do not detect any other influence on the photoelectron spectra~\cite{Schenk2010}.

The laser electric field $E(t)$ is modelled similarly to the definition given for the SMM, but with a Gaussian envelope. The wave function is propagated in time using the Crank-Nicolson method, and the resulting outgoing wavepacket is spectrally analyzed. \MK{Quantum mechanical re-scattering in the TDSE model can happen both at the potential step at the metal-vacuum interface and at the infinitely high potential wall confining the electronic wave function. We numerically observe that the main contribution to re-scattering stems from the infinitely high wall. This may lead to an overestimation of the re-scattered electrons as compared to the direct ones. However, here we are not interested in the direct part or the absolute yield of re-scattered electrons, hence this model effect is of no importance. }The peak electric field strength and the duration of the laser pulse were slightly adapted to $E_0 = 9.9\,\mathrm{GV/m}$ and 5.5\,fs, respectively, to match the experimentally observed spectrum in the plateau part. The peak electric field of $E_0 = 9.9\,\mathrm{GV/m}$ corresponds to a field enhancement factor of about \MK{6}. A comparison between experimental and simulation data is shown in Fig.~\ref{fig:ThTDSEExp}\TV{a}. The agreement is poor for the direct part but good for the plateau region. This likely results from the fact that a narrow one-dimensional potential well was chosen to model the metal, totally ignoring the vast number of de-localized electronic states and the band structure. For a full description a more elaborate model is needed. Also, plasmonic effects and other effects resulting from the 3-dimensionality cannot be simulated with this model. However, in the emission dynamics of wavepackets belonging to the plateau the evanescent (vacuum) part of the eigenstate wave functions plays the key role in the emission~\cite{Faisal2005} and is well described by the simple model. At the surface, states extend into the vacuum independent of their localized (surface state) or de-localized (bulk state) nature. This is in close analogy to the case of bound states in an atomic potential. Hence the potential well ground state used here can \OK{qualitatively} reflect the actual situation. Moreover, the dynamics of wavepackets belonging to the plateau is less sensitive to the exact shape of the surface potential due to the high kinetic energies gained after emission.

\begin{figure}
\includegraphics[width=0.9\columnwidth]{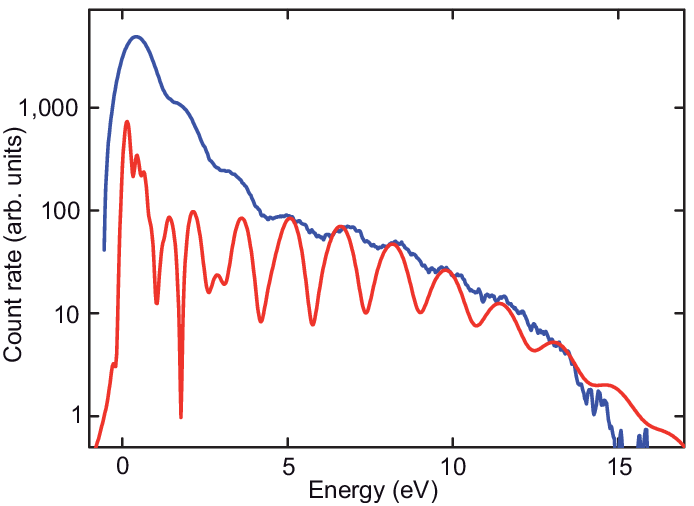}
\caption{{\bf Numerical integration of the time-dependent Schr{\" o}dinger equation.} \TV{{\bf a}, }Comparison of the numerical spectrum (red curve) with the experimental photoelectron spectrum (blue curve). Both spectra are averaged in the \mbox{C-E} phase. Good agreement is evident in the plateau, but the direct part is not well reproduced.\TV{{\bf b}, Initial state population (black curve) and laser electric field (red curve) of a 5.5-fs pulse with \mbox{C-E} phase $\phi_{\mathrm{CE}} = 0$ as a function of time. The population drops in a step-like manner (steps are marked with dashed horizontal lines), along with short-lived oscillations. The initial state is strongly depleted in two half-cycles of the pulse, where electric field strength reaches a maximum in the negative direction, thus pulling electrons out of the metal.}
}
\label{fig:ThTDSEExp}
\end{figure}

The analysis of the average peak visibility and the cut-off position depicted in Fig.~3d was carried out in the same way as for the experimental data. According to the SMM the highest cut-off position is found for a \mbox{C-E} phase of $\sim${}$0.03\pi$. The TDSE model, however, gives a value of about $-0.22\pi$. This discrepancy is probably due to the fact that the image potential force is neglected in the SMM. Its inclusion would cause a slight phase shift, in close analogy to Coulomb effects in above-threshold ionization of atomic gases~\cite{MilosevicBecker2006}. We chose to calibrate the experimental \mbox{C-E} phase offset according to the behaviour of the cut-off in the TDSE model.

\TV{[Paragraph on Fig. S4b dropped]}
\TV{In Fig.~\ref{fig:ThTDSEExp}b we show the ground state population as a function of time for a cosine-like pulse ($\phi_{\mathrm{CE}} = 0$). A step-like structure is visible, along with short-lived oscillations. This indicates that the emission process resolves the temporal structure of the laser electric field. Emission is strongly confined to instants in time when the laser electric field is negative and bends down the potential barrier. Moreover, it is highly nonlinear in the instantaneous electric field strength. These observations support the notion of an optically induced tunnelling process.}

\MS{
\section{Possible thermal effects}

The question of whether thermal emission of electrons may take place deserves consideration. Thermally-enhanced field emission from tips has been observed for continuous wave laser illumination~\cite{Lee1980} and for
pulsed laser illumination~\cite{Hilbert2009, Kealhofer2011}. A similar effect has been observed in laser-assisted scanning tunneling microscope measurements~\cite{Merschdorf2002} and in photoemission studies of metal nanoclusters with $80$\,fs IR pulses~\cite{Gloskovskii2007}. The transition from prompt multi-photon
photoemission to thermally enhanced emission depends on various parameters (e.g. pulse duration, fluence, geometrical dimensions) and is discussed in \bibpunct{}{}{,}{n}{}{}Ref.~\cite{Kealhofer2011}\bibpunct{}{}{,}{s}{}{} in great detail. We estimate the thermalized electron gas temperature for our parameters to around $2,000$\,K, which is much too small for pure thermal emission over the effective barrier height of about $5.2$\,eV (corresponding to $\sim 60,000$\,K). Hence we can exclude a significant contribution of thermally enhanced field emission. We experimentally confirmed this by measuring pump-probe (autocorrelation) signals in the electron current and did not observe any delayed
component, unlike what was demonstrated in \bibpunct{}{}{,}{n}{}{}Ref.~\cite{Fujimoto1984}\bibpunct{}{}{,}{s}{}{}. Note that within the time window of the laser (pulse)-tip interaction of $\sim$6\,fs an equilibrium temperature of the electron gas is not established. Hence, in experiments with comparable experimental parameters a strong non-equilibrium excited carrier distribution with a step-like structure has been observed~\cite{Ropers2007, Yanagisawa2011}.}

\section{Related work: Femtosecond-laser induced electron emission from metal tips}

The combination of femtosecond laser pulses and a sharp metal tip (tungsten or gold) has been shown to represent an ultrafast laser-driven electron emitter on the nanometre scale~\cite{Hommelhoff_PRL2006, Hommelhoff_PRL2006_2, Ropers2007, Barwick_Batelaan_NJP2007, Hilbert2009}. The system was used to demonstrate the non-dispersive nature of the Aharanov-Bohm effect under well-controlled experimental conditions~\cite{Caprez_Batelaan_PRL2007} and to implement a new microscopy technique for imaging nanostructures~\cite{Ropers2007}. Laser-illuminated tips were also explored for a prospective application as high-brightness electron source for free electron lasers~\cite{Ganter2008,Tsujino2008,Tsujino2009}. Tip geometry and laser-induced electron dynamics (surface waves) lead to a significant enhancement of the laser electric field at the tip's apex~\cite{Martin2001, Hommelhoff_PRL2006, Hommelhoff_PRL2006_2, Ropers2007}. This effect has been thoroughly investigated and can even be used to control the emission sites on the tip~\cite{Yanagisawa2009,Yanagisawa2010}. Very recently it was shown that it is possible to enter the strong-field regime of photoemission~\cite{Bormann2010,Schenk2010}.
